\documentclass[letterpaper, 10pt, conference]{ieeeconf}

\IEEEoverridecommandlockouts
\usepackage{amsmath}
\usepackage{amssymb}
\usepackage{graphicx}
\usepackage{subcaption}
\usepackage{booktabs}
\usepackage{siunitx}
\makeatletter
\let\NAT@parse\undefined
\makeatother
\usepackage{hyperref}
\hypersetup{
    colorlinks=false,
    pdfborder={0 0 1},
}
\usepackage{orcidlink}
\usepackage{cleveref}
\usepackage{xcolor}
\usepackage{tikz}
\usetikzlibrary{positioning, arrows.meta, fit, backgrounds, calc, decorations.pathreplacing, shapes.geometric}

\definecolor{colGNSS}{RGB}{0,114,178}
\definecolor{colUWB}{RGB}{0,158,115}
\definecolor{colWiFi}{RGB}{230,159,0}
\definecolor{colBLE}{RGB}{204,121,167}
\definecolor{col5G}{RGB}{213,94,0}

\definecolor{zoneOut}{RGB}{26,120,120}
\definecolor{zoneTrans}{RGB}{90,90,90}
\definecolor{zoneIn}{RGB}{150,30,60}

\title{\LARGE \bf
Bridging the Indoor-Outdoor Gap:\\Cross-Technology Ranging for Seamless Robot Navigation
}

\author{Paul Schwarzbach\,\orcidlink{0000-0002-1091-782X}$^{1}$%
\thanks{$^{1}$Paul Schwarzbach is with the Institute of Traffic Telematics,
        TUD Dresden University of Technology, 01069 Dresden, Germany
        {\tt\small paul.schwarzbach@tu-dresden.de}}%
}

\begin{document}

\maketitle
\thispagestyle{empty}
\pagestyle{empty}

\begin{abstract}

Mobile robots that move between outdoor and indoor environments still struggle with consistent positioning. Satellite-based and terrestrial ranging each work well in their home domains, but combining them at the raw measurement level has received little attention, and the building boundary is precisely where both classes degrade. This paper reports preliminary observations from the HYMN dataset, which time-synchronizes raw measurements from GNSS, Ultra-Wideband (UWB), WiFi Fine Time Measurement (FTM), and Bluetooth Low Energy (BLE) against millimeter-level ground truth in an industrial setting. Per-zone measurement availability and ranging-residual behavior are characterised. The two technology classes turn out to be complementary, and the indoor-outdoor transition is where their weaknesses overlap. The dataset is publicly available.

\end{abstract}

\section{INTRODUCTION}

Mobile robots in logistics, inspection, and search-and-rescue routinely cross the boundary between outdoor and indoor environments. State estimation across this divide depends on a globally consistent position, but the available ranging infrastructure changes at the building wall. Outdoors, satellite-based Global Navigation Satellite Systems (GNSS) dominate; once the robot is inside, terrestrial systems such as Ultra-Wideband (UWB), WiFi, and Bluetooth Low Energy (BLE) take over.

Tightly-coupled raw-measurement fusion has been demonstrated for specific technology pairs, in particular GNSS+UWB~\cite{Guo_gnss_uwb_time_calibration_2023,Michler2025_PLANS}. Across the broader set of ranging modalities relevant to indoor-outdoor navigation, however, a unified measurement-level view of observability and error structure is still missing. The gap is felt most in degraded environments, where multipath, non-line-of-sight (NLOS) conditions, and signal interference dominate.

Radio-based positioning has matured within individual domains. GNSS draws on decades of raw pseudorange processing, error modeling, and integrity monitoring~\cite{Jin2024_Survey_GNSS_Cooperative_Positioning}. UWB, WiFi FTM, BLE, and 5G NR are each mature enough for operational deployment. These communities evolved in parallel, with their own terminology, incompatible standards, and no shared benchmark spanning multiple modalities~\cite{Schwarzbach2026_Access_TermFragmentation}. The mismatch becomes a practical problem at the building boundary, where a state estimator must either switch between different observation types or fuse them simultaneously without an agreed handover protocol~\cite{MallikMLforSIO2023}.

\begin{figure}[!t]
    \centering
    \resizebox{\columnwidth}{!}{%
    \begin{tikzpicture}[
        font=\sffamily\footnotesize,
        >=Stealth,
        paneltitle/.style={font=\sffamily\footnotesize\bfseries, text=black!70, anchor=base, inner sep=1pt},
        arrow/.style={->, line width=1.1pt, black!55},
        techlabel/.style={font=\sffamily\scriptsize\bfseries, inner sep=1pt},
    ]

    \begin{scope}[local bounding box=panelA,
        techbox/.style={draw, line width=0.6pt, rounded corners=2pt,
                        minimum width=1.8cm, minimum height=0.50cm,
                        font=\sffamily\footnotesize\bfseries,
                        align=center, inner sep=2pt}]
        \node[techbox, draw=colGNSS, fill=colGNSS!8, text=colGNSS]          (gnss) at (1.00, 2.085) {GNSS};
        \node[techbox, draw=colUWB,  fill=colUWB!8,  text=colUWB,  below=3pt of gnss] (uwb)  {UWB};
        \node[techbox, draw=colWiFi, fill=colWiFi!8, text=colWiFi, below=3pt of uwb]  (wifi) {WiFi};
        \node[techbox, draw=colBLE,  fill=colBLE!8,  text=colBLE,  below=3pt of wifi] (ble)  {BLE};

        \path (0, 0) rectangle (2.00, 2.50);
    \end{scope}

    \begin{scope}[shift={($(panelA.east)+(1.78, -0.83)$)}, local bounding box=panelB, xscale=0.055, yscale=0.055]
        \draw[thin, black!35, rounded corners=2pt] (-21.5, -7.2) rectangle (12.8, 37.4);

        \foreach \x/\y in {%
            4.104/32.903, 1.141/29.928, -1.866/32.92, -4.86/29.889, -7.839/32.869,
            -10.831/29.882, -7.922/26.872, -10.792/23.891, -4.86/23.891, -1.907/26.909,
            1.136/23.935, 4.089/26.995, 3.827/20.947, 1.109/17.981, -1.841/20.912,
            -4.884/17.95, -7.842/20.916, -10.893/17.915, 4.083/11.839, 1.146/8.945,
            -1.916/11.94, -4.87/8.975, -7.915/11.927, -10.875/9.058, -10.872/3.01,
            -7.895/5.977, -4.875/2.989, -1.882/5.995, 1.149/2.955, 4.155/5.798,
            4.096/-0.04, -1.859/-0.04, -7.871/-0.026, 1.118/14.944, -4.892/14.912,
            -10.902/14.812, -20.14/-3.382, -17.467/-3.033, -14.809/-3.049,
            -10.833/-3.07, -7.814/-3.057, -4.836/-3.078, -1.841/-3.097, 1.175/-3.083,
            4.162/-3.074, 6.813/-2.964, 9.365/-2.779, 12.056/-2.534%
        }{
            \node[circle, fill=black!45, inner sep=0pt, minimum size=1.2pt] at (\x,\y) {};
        }

        \foreach \x/\y in {%
            3.685/22.496, 4.059/6.256, -3.969/-6.381, -12.157/7.832, -11.976/24.21,
            -3.651/32.95, 4.754/35.133, 2.035/-6.415, -9.984/-6.38, -12.099/36.584%
        }{
            \node[rectangle, fill=colUWB, inner sep=0pt, minimum size=3.0pt] at (\x,\y) {};
        }

        \foreach \x/\y in {%
            3.69/22.509, 4.058/6.255, -3.964/-6.37, -12.172/7.83, -11.984/24.218%
        }{
            \node[circle, fill=colBLE, inner sep=0pt, minimum size=2.4pt] at ($(\x,\y)+(1.8,1.8)$) {};
        }

        \foreach \x/\y in {%
            3.644/22.526, 4.012/6.247, -3.988/-6.327, -12.104/7.831, -11.927/24.211,
            -3.673/32.893%
        }{
            \node[regular polygon, regular polygon sides=3, fill=colWiFi, inner sep=0pt, minimum size=2.8pt] at ($(\x,\y)+(-1.8,-1.8)$) {};
        }
    \end{scope}

    \begin{scope}[shift={($(panelB.east)+(0.80, -0.85)$)}, local bounding box=panelC]
        \draw[->, black!55, line width=0.5pt, >=Stealth] (0, 0) -- (0, 2.10);
        \draw[->, black!55, line width=0.5pt, >=Stealth] (0, 0) -- (2.25, 0);
        \draw[black!25, dashed, line width=0.3pt] (0, 1.70) -- (2.20, 1.70);

        \node[font=\sffamily\scriptsize, text=black!60, anchor=east, inner sep=1pt] at (-0.03, 1.70) {1};
        \node[font=\sffamily\scriptsize, text=black!60, anchor=east, inner sep=1pt] at (-0.03, 0.00) {0};
        \node[font=\sffamily\scriptsize, text=black!70, anchor=north] at (1.10, -0.04) {ranging residual};

        \draw[zoneOut, line width=0.7pt]
            plot[smooth, tension=0.55] coordinates
            {(0,0) (0.13, 0.81) (0.30, 1.36) (0.55, 1.62) (0.93, 1.69) (2.17, 1.70)};

        \draw[zoneTrans, line width=0.7pt, dashed]
            plot[smooth, tension=0.55] coordinates
            {(0,0) (0.17, 0.38) (0.47, 0.98) (0.89, 1.49) (1.40, 1.67) (2.17, 1.70)};

        \draw[zoneIn, line width=0.7pt, dash pattern=on 2pt off 1pt on 0.5pt off 1pt]
            plot[smooth, tension=0.55] coordinates
            {(0,0) (0.30, 0.17) (0.72, 0.60) (1.23, 1.15) (1.78, 1.53) (2.17, 1.67)};

        \node[font=\sffamily\scriptsize, text=zoneOut,   anchor=west] at (1.32, 0.78) {outdoor};
        \node[font=\sffamily\scriptsize, text=zoneTrans, anchor=west] at (1.32, 0.53) {transition};
        \node[font=\sffamily\scriptsize, text=zoneIn,    anchor=west] at (1.32, 0.28) {indoor};

        \draw[zoneOut,   line width=0.7pt]                                        (1.18, 0.78) -- (1.30, 0.78);
        \draw[zoneTrans, line width=0.7pt, dashed]                                (1.18, 0.53) -- (1.30, 0.53);
        \draw[zoneIn,    line width=0.7pt, dash pattern=on 2pt off 1pt on 0.5pt off 1pt] (1.18, 0.28) -- (1.30, 0.28);

        \path (0, 0) rectangle (2.60, 2.50);
    \end{scope}

    \coordinate (titleY) at (0, 2.60);
    \node[paneltitle] at (panelA.north |- titleY) {Ranging Techn.};
    \node[paneltitle] at (panelB.north |- titleY) {HYMN Dataset};
    \node[paneltitle] at (panelC.north |- titleY) {Evaluation};

    \coordinate (arrowY) at (0, 1.25);
    \draw[arrow] ([xshift=2pt]panelA.east |- arrowY) -- ([xshift=-2pt]panelB.west |- arrowY);
    \draw[arrow] ([xshift=2pt]panelB.east |- arrowY) -- ([xshift=-2pt]panelC.west |- arrowY);

    \end{tikzpicture}%
    }
    \caption{Four ranging technologies (left), recorded by the HYMN measurement campaign (middle), characterized per zone to inform robotic state estimation (right).}
    \label{fig:flowgraph}
\end{figure}
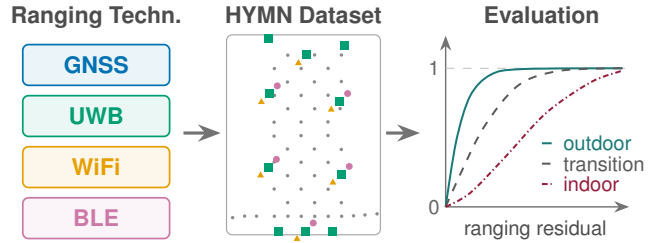

\begin{figure}[htpb]
    \centering
    \includegraphics[width=\columnwidth]{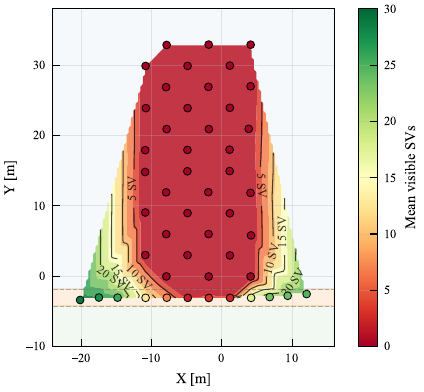}
    \caption{Mean number of visible GNSS satellites across the dataset. Dashed bands mark the transition zone.}
    \label{fig:motivation}
\end{figure}

\Cref{fig:motivation} shows what this looks like in practice. The mean number of visible GNSS satellites drops from more than 20 outdoor to zero inside the hall, with a narrow transition strip at the gate where satellite visibility is already marginal. Satellite-only positioning fails precisely in the region a robot must cross, so complementary terrestrial ranging is a requirement rather than an option.

Several recent works approach this gap from different sides. Multi-technology datasets now support cross-system comparison~\cite{Ammad2026_HYMN_Descriptor}, and early fusion studies have shown that heterogeneous ranging sources can be combined for seamless positioning~\cite{Michler2025_PLANS}. What is still missing is a systematic picture of how raw ranging observations from different technologies behave {relative to each other} across spatial zones. Principled error models for integrity-aware multi-technology state estimation are equally scarce.

This paper presents a preliminary assessment of cross-technology ranging measurements from the HYMN (HYbrid Multi-technology Navigation) dataset, outlined visually in \Cref{fig:flowgraph}. The contributions are:
\begin{itemize}
    \item An empirical per-zone (outdoor, transition, indoor) characterization of four time-aligned ranging technologies (GNSS, UWB, WiFi FTM, BLE) against millimeter-level ground truth, looking at both measurement availability and residual behavior. The indoor-outdoor transition turns out to be the regime in which satellite and terrestrial systems degrade at the same time.
    \item A measurement-level framing of this heterogeneous data aimed at robotic state estimator design, with the goal of connecting the GNSS-navigation and indoor-positioning communities.
\end{itemize}

\section{MEASUREMENTS AND RANGING MODEL}

\subsection{HYMN Measurement Campaign}

The observations come from the HYMN dataset~\cite{Ammad2026_HYMN_Descriptor}, recorded at an industrial facility in Torgau, Germany. A \SI{44}{\meter} $\times$ \SI{18}{\meter} hall with a wide driveway gate creates a natural indoor-outdoor transition, and a grid of measurement points covers the outdoor, transition, and indoor zones. Four ranging technologies operated simultaneously: GNSS (NovAtel PwrPak7), UWB (ZigPos RTLS), WiFi FTM (HPE Aruba 630), and BLE (Metirionic DMK-215). A Leica TS16 total station provides ground truth at millimeter precision and fixes the local Cartesian frame. Through surveyed tie points, GNSS ECEF observations project into that frame, so all four technologies share a common reference. Per-system timestamps are aligned to a shared timescale for direct comparison. The dataset is publicly available~\cite{Ammad2026_HYMN_dataset}.

\begin{figure}[thbp]
    \centering
    \includegraphics[width=0.78\columnwidth]{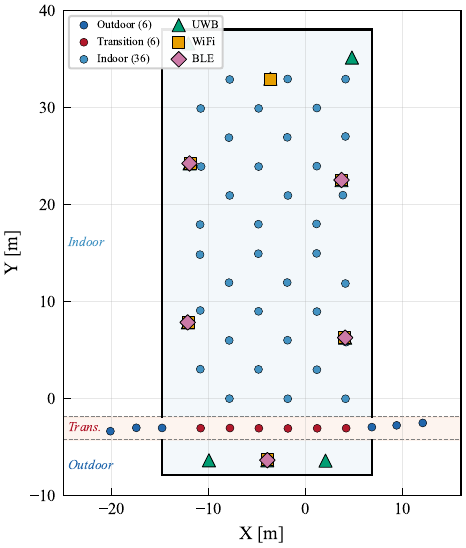}
    \caption{Campaign layout: industrial hall with driveway, zone partitioning, measurement-point grid, and anchor positions.}
    \label{fig:layout}
\end{figure}

\Cref{fig:layout} shows the physical site. An autonomous platform operating in a warehouse or industrial facility has to handle exactly this kind of transition between covered and open areas, with varying satellite visibility and anchor coverage.

\subsection{Ranging Residual Model}

The four HYMN technologies rest on different physical principles but produce observations that fit a shared three-term structure:
\begin{equation}
    \rho_i^{(s)} = d_i^{(s)} + b^{(s)} + \eta_i^{(s)} + \epsilon_i^{(s)}
    \label{eq:observation_model}
\end{equation}
where $\rho_i^{(s)}$ is the raw range observation from system $s$ to reference point $i$ and $d_i^{(s)}$ is the true geometric distance. The three error terms behave qualitatively differently. The deterministic bias $b^{(s)}$ covers clock offsets, modelled atmospheric delays, and per-anchor calibration offsets, and can be corrected or removed by differencing. The environment-dependent stochastic error $\eta_i^{(s)}$ captures NLOS excursions and multipath. It is positive in expectation, state-dependent, and typically heavy-tailed, which rules out Gaussian handling and calls for robust estimation or non-parametric measurement models. The zero-mean measurement noise $\epsilon_i^{(s)}$ absorbs whatever remains. Since $d_i^{(s)}$ is known from the total-station survey, the residual
\begin{equation}
    r_i^{(s)} = \rho_i^{(s)} - d_i^{(s)} = b^{(s)} + \eta_i^{(s)} + \epsilon_i^{(s)}
    \label{eq:residual}
\end{equation}
is directly accessible. The per-technology residual budget differs in which term dominates. For a robotic state estimator, these technology-specific characteristics determine measurement noise covariance, outlier rejection thresholds, and appropriate weighting in a fusion filter.

\textbf{GNSS Pseudoranges} are one-way time-of-flight observations corrected with precise satellite clock and orbit products and the standard single-frequency error terms (ionosphere, troposphere, Earth-rotation). Between-Satellite Single Differencing against the highest-elevation pivot then removes the receiver clock and residual common-mode atmospheric and orbit errors. What remains in $\eta$ is dominated by NLOS reception near the hall gate and by multipath, which broaden the residuals.

\textbf{UWB Two-Way Ranging} uses double-sided exchanges to cancel most of the crystal offset between tag and anchor. Factory antenna-delay calibration absorbs the residual constant term, leaving a LOS residual well approximated by a zero-mean Gaussian with centimetre-scale $\epsilon$~\cite{Schwarzbach2022_UWB_Statistical}. The NLOS regime breaks that model. First-path detection bias under wall and clutter penetration injects sparse, positive-mean excursions into $\eta$, producing the heavy-tailed outlier groups visible at longer ranges in the Torgau data and motivating the mixture-model treatment in~\cite{Schwarzbach2022_UWB_Statistical}.

\textbf{WiFi Fine Time Measurement} is the WiFi analogue of UWB TWR. A session of 802.11mc timestamp exchanges runs between initiator and responder, and $\mathrm{RTT} = (t_4-t_1) - (t_3-t_2)$ cancels the inter-device clock offset on the same round-trip principle, pushing $b$ down to per-anchor cable and antenna delay~\cite{KosekSzott2026_WiFi_FTM}. Time-of-flight resolution is set by the channel bandwidth (20--80~MHz on the deployed Aruba 630), keeping $\epsilon$ above the UWB noise floor. Multipath and NLOS reflections fill $\eta$ and bias the first-arrival timing.

\textbf{BLE Phase-Based Ranging} on the Metirionic DMK-215 sweeps the 2.4~GHz ISM band and collects IQ samples at discrete channel frequencies. An inverse Fourier transform reconstructs a coarse channel impulse response, from which range follows either from the first peak or from the phase slope across frequency~\cite{Suresh2025_BLE_PBR}. Here $b$ and $\eta$ do not decouple cleanly. Phase-calibration error, IQ imbalance, and ISM-band multipath fold into the residual, yielding heavy tails.

\section{CROSS-TECHNOLOGY OBSERVATIONS}

\subsection{System Availability Across Zones}

\begin{figure*}[t]
    \centering
    \includegraphics[width=1\textwidth]{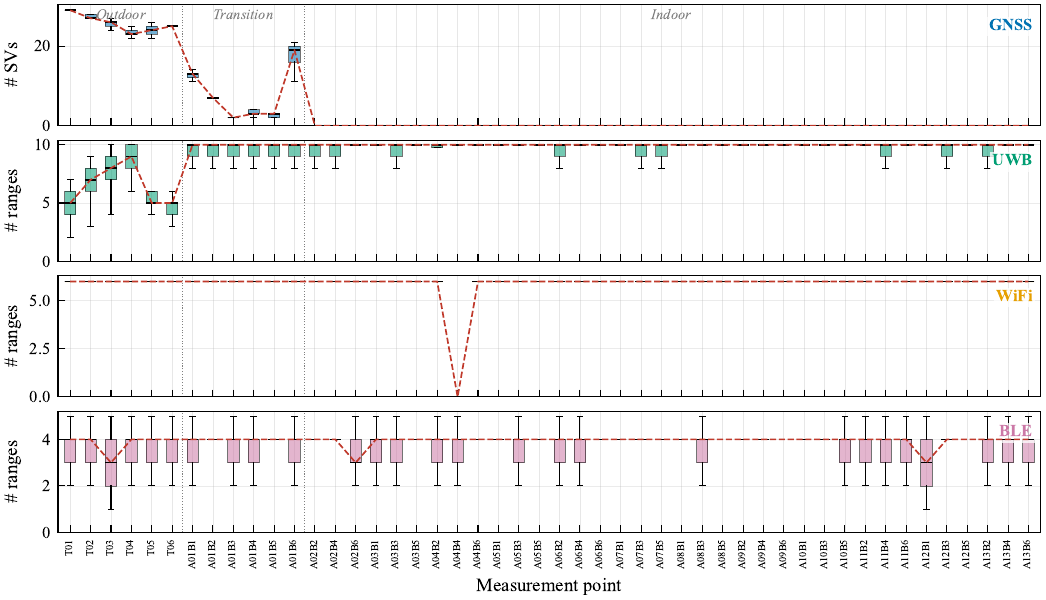}
    \caption{Per-point observation counts: valid ranges per epoch (UWB, WiFi, BLE) or visible satellites (GNSS).}
    \label{fig:availability}
\end{figure*}

\Cref{fig:availability} reports the per-epoch count of simultaneous ranging observations (UWB, WiFi, BLE) or visible satellites (GNSS) at each of the 48 measurement points.

\begin{figure*}[th]
    \centering
    \includegraphics[width=.92\textwidth]{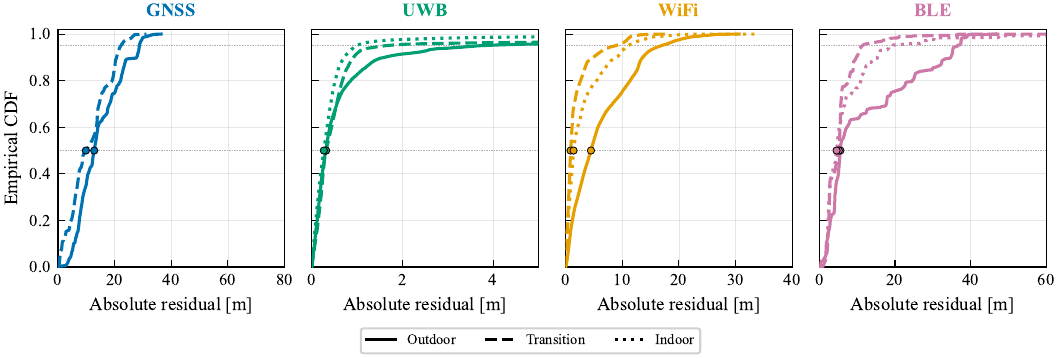}
    \caption{Empirical CDFs of absolute ranging residuals (independent x-axes). Linestyles encode zones and dots mark medians.}
    \label{fig:cdf}
\end{figure*}

\begin{figure*}[th]
    \centering
    \begin{subfigure}[b]{0.25\textwidth}
        \centering
        \includegraphics[width=\linewidth]{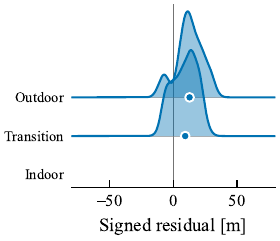}
        \caption{GNSS}
        \label{fig:ridgeline_gnss}
    \end{subfigure}
    \hfill
    \begin{subfigure}[b]{0.21\textwidth}
        \centering
        \includegraphics[width=\linewidth]{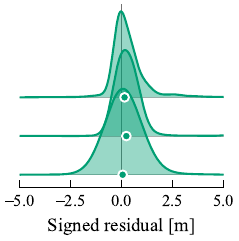}
        \caption{UWB}
        \label{fig:ridgeline_uwb}
    \end{subfigure}
    \hfill
    \begin{subfigure}[b]{0.21\textwidth}
        \centering
        \includegraphics[width=\linewidth]{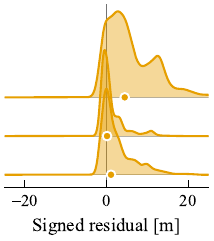}
        \caption{WiFi}
        \label{fig:ridgeline_wifi}
    \end{subfigure}
    \hfill
    \begin{subfigure}[b]{0.21\textwidth}
        \centering
        \includegraphics[width=\linewidth]{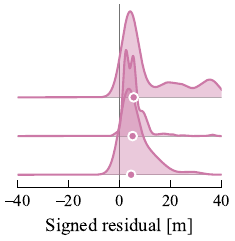}
        \caption{BLE}
        \label{fig:ridgeline_ble}
    \end{subfigure}
    \caption{Kernel-density of ranging residuals per technology and zone (independent x-axes). Dots mark medians.}
    \label{fig:ridgeline}
\end{figure*}

Three patterns stand out. First, GNSS satellite counts drop from 25--30 SVs outdoors to zero across nearly all indoor points, with the transition zone showing a sharp gradient: points adjacent to the open gate still see 12--20 SVs, while points only a few meters deeper into the hall collapse to a handful. Second, UWB range counts {increase} from roughly five observations outdoors to the full anchor complement (10) indoors. This is caused by the anchor layout, where indoor points lie inside the UWB network while outdoor points see only the subset of anchors with direct line of sight through the gate. Third, WiFi and BLE remain broadly available across all zones with point-level variability dominated by anchor geometry and obstruction rather than by zone membership. The isolated WiFi drop at A04B4 reflects a single logged epoch with no successful FTM exchange and is a data-capture artifact rather than a coverage phenomenon.

\subsection{Ranging Residual Characteristics}

\Cref{tab:residuals} summarises the residual distribution from \Cref{eq:residual} per technology and per zone. The reported metrics are the mean residual (Bias), the median, the standard deviation around the mean ($\sigma$), and the 95\textsuperscript{th} percentile of the absolute residual ($P_{95}|e|$). All four technologies produce predominantly positive residuals, in line with $\eta$ being positive in expectation. Observation counts also differ markedly across zones, which inflates $\sigma$ in the larger zones through a higher chance of capturing rare extreme values. Hence, $P_{95}|e|$ is the more robust cross-zone descriptor.

\begin{table}[t]
\centering
\caption{Signed ranging residual statistics per technology and zone. $^\dagger$GNSS: pivot BSSD against the highest-elevation satellite per epoch, pooled over GPS~L1 and Galileo~E1 pseudoranges. Pseudoranges are corrected for satellite clock (CODE precise), Klobuchar ionosphere (broadcast coefficients, GPS Klobuchar applied as single-frequency proxy to E1), simple exponential troposphere, and Sagnac/Earth-rotation before differencing; orbits from CODE precise SP3.}
\label{tab:residuals}
\setlength{\tabcolsep}{3pt}
\begin{tabular}{ll r r r r r}
\toprule
Technology & Zone & Bias & Med. & $\sigma$ & RMSE & $P_{95}|e|$ \\
 & & {[\si{\meter}]} & {[\si{\meter}]} & {[\si{\meter}]} & {[\si{\meter}]} & {[\si{\meter}]} \\
\midrule
GNSS$^\dagger$ & Outdoor & +12.61 & +12.79 & 10.16 & 16.19 & 28.73 \\
 & Transition & +8.94 & +9.37 & 9.46 & 13.02 & 22.26 \\
 & Indoor & +11.06 & +8.93 & 10.52 & 15.23 & 32.97 \\
\midrule
UWB & Outdoor & +0.75 & +0.14 & 3.15 & 3.24 & 3.70 \\
 & Transition & +0.25 & +0.23 & 4.46 & 4.47 & 1.62 \\
 & Indoor & +0.68 & +0.05 & 9.97 & 10.00 & 0.99 \\
\midrule
WiFi & Outdoor & +5.94 & +4.47 & 5.75 & 8.27 & 16.87 \\
 & Transition & +1.17 & +0.13 & 2.98 & 3.20 & 8.82 \\
 & Indoor & +2.62 & +1.10 & 4.09 & 4.86 & 11.00 \\
\midrule
BLE & Outdoor & +12.06 & +5.54 & 11.92 & 16.95 & 36.56 \\
 & Transition & +5.48 & +5.12 & 4.60 & 7.15 & 11.33 \\
 & Indoor & +7.71 & +4.53 & 13.83 & 15.83 & 19.37 \\
\bottomrule
\end{tabular}
\end{table}

\textbf{UWB} stays unbiased on the LOS bulk in every zone, with median residuals near zero (indoor \SI{0.05}{\meter}). The bulk distribution is tightest indoors: $P_{95}|e| = \SI{0.99}{\meter}$, against \SI{1.62}{\meter} in the transition zone and \SI{3.70}{\meter} outdoor. The large indoor $\sigma \approx \SI{10}{\meter}$ is not a bulk property. It is inflated by gross outliers potentially caused by target confusions~\cite{Schwarzbach2022_UWB_Statistical}, which the larger indoor sample size is more likely to surface. With observation counts unequal across zones, $\sigma$ is best read per-zone rather than as a cross-zone comparison.

\textbf{WiFi FTM} carries a positive RTT-calibration offset that varies by zone: the transition zone is tightest (Bias \SI{1.17}{\meter}, $\sigma$ \SI{2.98}{\meter}), the indoor zone intermediate (Bias \SI{2.62}{\meter}, $\sigma$ \SI{4.09}{\meter}), and the outdoor zone widest (Bias \SI{5.94}{\meter}, $\sigma$ \SI{5.75}{\meter}). The non-monotonic ordering is a per-zone signature to which anchor geometry, range distribution, and obstruction patterns plausibly all contribute.

\textbf{BLE PBR} carries a large positive bias in every zone (\SIrange{5}{12}{\meter}) and the widest spread of the four systems (outdoor $\sigma \approx \SI{12}{\meter}$). The offset is consistent with the phase-based pipeline described in Section~II: residual phase-calibration error, IQ imbalance, and ISM-band multipath fold into the first-peak and phase-slope estimates in the same direction rather than cancelling. The Bluetooth standard has since moved on, with BLE Channel Sounding combining phase-based ranging and round-trip-time measurements~\cite{Suresh2025_BLE_PBR}, so the results here characterise the deployed Metirionic DMK-215 rather than current state-of-the-art BLE ranging.

\textbf{GNSS BSSD} carries a residual positive bias of \SIrange{9}{13}{\meter} and $\sigma \approx \SI{10}{\meter}$ across all three zones, consistent with single-frequency model residuals. The indoor median (\SI{8.9}{\meter}) tracks the outdoor median (\SI{12.8}{\meter}), though the indoor sample reflects only the handful of satellites that penetrate the gate at favourable elevations, biasing the surviving residuals toward the easy cases.

The four technologies leave qualitatively different residual signatures. UWB pairs a near-zero LOS bulk with sparse, heavy-tailed NLOS excursions, while WiFi FTM and BLE PBR carry zone-dependent positive biases on top of moderate spread. GNSS BSSD sits closer to the terrestrial systems in spread, with a zone-stable bias of \SIrange{9}{13}{\meter} and $\sigma \approx \SI{10}{\meter}$ dominated by single-frequency model residuals. Zone signatures themselves differ: UWB spread inflates indoors, WiFi tightens in the transition zone, BLE stays broad everywhere, and GNSS spread is comparable across zones while availability collapses. Residuals also vary spatially within zones with anchor geometry and obstruction.

\section{DISCUSSION AND OUTLOOK}

\subsection{Implications for Robot Navigation}

\textbf{The transition zone is the critical region.} Satellite and terrestrial ranging degrade together at the building boundary. GNSS satellite visibility drops sharply (\Cref{fig:motivation}), dilution of precision grows, and multipath susceptibility rises. Only a subset of UWB, WiFi, and BLE anchors remain visible through the gate, and at an unfavourable geometry. The handover between the two system classes happens where both have degraded, so a state estimator that trusts measurements equally across zones will fail where it matters most. Maintaining positioning quality through the transition therefore calls for zone-aware measurement noise adaptation, or integrity-monitoring-based measurement selection.

\textbf{Beyond zero-mean Gaussian noise.} The residual signatures of Section~III rule out a single covariance per technology. UWB shows a near-zero bulk with sparse heavy-tailed NLOS excursions, WiFi FTM shows zone-dependent positive biases on top of moderate spread, BLE PBR shows large positive biases and the widest spread of the four systems, while GNSS BSSD holds a moderate zone-stable bias of \SIrange{9}{13}{\meter} with $\sigma \approx \SI{10}{\meter}$ regardless of zone. Two estimator-design families can absorb that structure. Bayesian filters with non-parametric measurement representations carry the full residual distribution into the update step~\cite{Michler2025_PLANS}. Robust factor-graph back-ends reach the same goal on the optimisation side, down-weighting outliers through M-estimators, switchable constraints, or dynamic covariance scaling. Which of the two is preferable is a question of the surrounding estimator stack, not of the residual evidence on its own.

\textbf{Self-calibrating infrastructure.} Per-anchor calibration enters the observation model of \Cref{eq:observation_model} through the deterministic bias term $b^{(s)}$. Manual surveying of every anchor remains the dominant practice and limits scalability for robots that move between sites. Recent work on collaborative self-calibration shows that sub-metre anchor positioning is reachable without dedicated surveying~\cite{Jung2026_CollabSelfCalib_submitted}, collapsing the deployment cost of terrestrial ranging to roughly that of dropping anchors in place.

\subsection{Bridging Communities}

\textbf{What each community brings.} The GNSS community has built a long tradition of rigorous per-source error budgets, atmospheric and ephemeris modelling, ambiguity resolution, and integrity monitoring, with RAIM as one well-known instance. On the indoor-positioning side the strengths lie elsewhere: a heterogeneous set of ranging modalities, environment-aware empirical error models, and self-calibrating infrastructure that handles deployment uncertainty. Each tradition addresses one half of the indoor-outdoor problem in depth, but neither has historically been forced to operate across the boundary.

\textbf{What is already transferring.} Robust factor-graph optimisation, originally developed in robotics and SLAM for outlier-prone landmark observations, is now being adopted for urban-canyon GNSS where multipath-corrupted pseudoranges break Gaussian filters. In the other direction, integrity and protection-level concepts from GNSS are starting to appear in safety-critical indoor robotics, where formal guarantees on positioning error become a deployment prerequisite. The exchange is held back less by the technical content than by language: identical operations carry different names in each community~\cite{Schwarzbach2026_Access_TermFragmentation}, which adds an avoidable barrier on top of the substantive ones.

\subsection{Limitations}

The observations presented here are preliminary. All measurements were collected at static points, not along continuous robot trajectories. The number of measurement points covers all three zones but remains modest, and the observation counts are unevenly distributed across them; balanced sampling is on the agenda for follow-up campaigns. Error model fitting (parametric distributions, zone-dependent parameterization, cross-technology correlation analysis) is ongoing and will be reported in subsequent publications. 

\section*{Declaration on Generative AI}
During the preparation of this work, the author used Anthropic's Claude (via the Claude Code CLI) to assist with drafting and editing of the manuscript as well as with development of the evaluation code. All AI-generated content was reviewed, verified, and edited by the author, who takes full responsibility for the final content of this publication.


\bibliographystyle{IEEEtran}
\bibliography{references}

\end{document}